\documentclass[aps,prl,reprint,groupedaddress]{revtex4-2}
\usepackage{graphicx}
\usepackage{amstext} 
\usepackage{array}   
\newcolumntype{L}{>{$}l<{$}} 
\usepackage{pstricks}
\usepackage{color}
\usepackage{placeins}
\usepackage{amsmath}

\begin{document}


\title{Induced spontaneous symmetry breaking chain}

\author{Eduard~Boos}
\email{boos@theory.sinp.msu.ru, Eduard.Boos@cern.ch}
\affiliation{Skobeltsyn Institute of Nuclear Physics, Lomonosov Moscow State University, 119991, Moscow,Russia}



\begin{abstract}
The article discusses a scenario based on the idea of induced spontaneous symmetry breaking. In this type of scenario, spontaneous symmetry breaking is assumed at some highest energy level, which leads to a chain of several subsequent induced symmetry violations at lower levels caused by small mixtures between the levels. We present a simple model in which the idea is realized by small mixing between the levels using scalar portals. In this approach, the large difference between energy scales, for example the Planck scale and the electroweak scale, occurs due to  the product of several small factors proportional to the mixing coefficients. Dark matter fields can be formed in this scenario from matter fields on one or possibly several intermediate levels between the highest one and lowest one. 
\end{abstract}

\maketitle

\section{Introduction}
\label{intro}
The Standard Model, quantum gauge field theory with spontaneous breaking of electroweak symmetry, reached its logical conclusion with the discovery of the Higgs boson at the LHC \cite{Aad:2012tfa, Chatrchyan:2012ufa}. 
The successes of the SM in describing the processes of particle production and decay in terrestrial and space experiments are well known. However, the SM is unable to explain a number of observed phenomena in nature, such as baryon and lepton asymmetries in the Universe, dark energy and dark matter, small cosmological constant, mass and oscillations of neutrinos, etc. 

There are some internal problems of the SM, such as the problem of the negative mass parameter
 $\mu^2 =-|\mu^2|$, artificially introduced into the theory, the hierarchy problem associated with this parameter and the problem of electroweak scale stabilization on a value of about 100 GeV. 
 The negative parameter $\mu^2$ is necessary for the realization of the mechanism of spontaneous electroweak symmetry breaking (mechanism BEH \cite{Higgs:1964pj, Englert:1964et, Guralnik:1964eu}), but its origin as well as the magnitude of the corresponding EW scale of the order of 100 GeV remain unclear. 

Before the launch of the LHC, there was the so-called No-Lose-Theorem
\cite{Lee:1977yc, Lee:1977eg, Chanowitz:1986ar, Dicus:1992vj}, which, based on an analysis of the behavior of the scattering amplitudes of longitudinal modes of massive electroweak SM bosons and the requirement of perturbative unitarity, asserted that only two situations were possible: either the existence of a sufficiently light Higgs boson with a mass smaller $\sim$700~GeV, or a change of regime and the appearance of new physics on a scale of the order of 1 TeV. The Higgs boson with a mass of about 125 GeV has been discovered, and, accordingly, the argument about the magnitude of the new scale is gone. Today we do not know, either the closest to the SM scale of new physics, or, of course, what new physics is. The results of the search for manifestations of new physics at the LHC are still negative, although these negative results themselves are extremely important, narrowing the range of parameters or closing some possible variants for new physics. 

We are well aware that there is at least one scale that is significantly larger than the electroweak scale. This is the Planck scale $ \sim 10^{19}$ GeV. Note that many extensions of the Standard Model predict the presence of other intermediate scales. Grand Unification models, their supersymmetric or string motivated generalizations, predict the scale$ \sim 10^{15} \div 10^{16}$GeV, on which the running electroweak and strong coupling constants become equal or close. In models of neutrino physics, in particular, based on the see-saw mechanism, the scale $ \sim 10^{11} \div 10^{12}$GeV appears. In supersymmetric models or in models with extra space-time dimensions, scales of the order of 10 TeV are quite admissible. However, the answer to the question of exactly what and how phase transitions occur as we move from large to smaller scales remains very unclear. 

This short letter proposes a simple scenario based on the idea of induced spontaneous symmetry breaking. 
In this type of scenario, a chain of phase transitions can be realized. Only in the theory at the highest scale, it can be the Planck scale, there is a negative mass parameter $\mu^2$. Possibly the negative sign of this parameter is  due to nonlinear effects of quantum gravity. At this highest scale, a phase transition occurs, and this transition through mixing causes subsequent transitions in a chain in which one link (i.e. theory at a given scale) connects to an adjacent link (i.e. theory at the nearest scale) through a portal with a rather small mixing. Small mixing parameters between the two nearest links (theories on the nearest scales) in the chain lead to the fact that the farther from the lowest link a certain link in the chain is located, the weaker the hypothetical particles of the theory at this intermediate scale interact with the particles of the theory at the lowest level.  In this scenario, the scale of the highest level, say the Plank scale $M_{Pl}$ and the scale of the lowest level, say the electroweak (EW) scale of the SM are related by a coefficient proportional to the product of small mixing parameters between theories at intermediate levels (links in the chain). 

The next section presents a simple model based on scalar portals. Models with scalar portals, being the simplest portal extensions of the SM, allow one to describe dark matter (DM)  without contradicting existing data and theoretical bounds in certain areas of the parameter space of these models \cite{Silveira:1985rk, McDonald:1993ex, Burgess:2000yq, Davoudiasl:2004be,Schabinger:2005ei, Barbieri:2005ri,Patt:2006fw,Strassler:2006ri, Bowen:2007ia, Barger:2007im, Barger:2008jx, He:2008qm, Bock:2010nz, Englert:2011yb, Lebedev:2011aq, Mambrini:2011ik, Belanger:2012zr, Cline:2013gha, Gabrielli:2013hma, Falkowski:2015iwa, Robens:2015gla, Martin-Lozano:2015dja, Casas:2017jjg, Athron:2017kgt, Arcadi:2019lka, Kannike:2020qtw, Saez:2021oxl}. For a detailed overview of the Higgs portal model in connection with DM, including various experimental restrictions, see \cite{Arcadi:2019lka}. Higgs and scalar portals, their DM and cosmological implications are summarized in the recent review
 \cite{Lebedev:2021xey}. In the presented scenario with a chain of scalar portal models with spontaneously induced symmetry breaking, there can obviously be candidates for dark matter particles, possibly from different levels of the chain, due to the larger number of parameters in hand that are usually discussed. 

\section{Simple model}
\label{model}
Consider the Lagrangian of a model that describes a chain of complex scalar fields with mixing between the nearest neighbors:

\begin{eqnarray}
\label{model_lgrn}
L =  (D^{(1)}_{\nu} H_1)^\dag (D^{(1)\nu}H_1)  
- \mu_1^2 H_1^\dag H_1  
- \lambda_1 (H_1^\dag H_1)^2 \,\,\,\,\,\,\,\,& \\ \nonumber
+ \, L_{Fields(1)}
+ \, k_{12} (H_1^\dag H_1) (H_2^\dag H_2) &\\ \nonumber
+  (D^{(2)}_{\nu} H_2)^\dag (D^{(2)\nu}H_2) 
- \mu_2^2 H_2^\dag H_2 
- \lambda_2 (H_2^\dag H_2)^2 \\ \nonumber
+ \, L_{Fields(2)}
+ \, k_{23} (H_2^\dag H_2) (H_3^\dag H_3) \\ \nonumber
 \dots \\ \nonumber
+  (D^{(N-1)}_{\nu} H_{N-1})^\dag (D^{(N-1)\nu}H_{N-1}) 
- \mu_{N-1}^2 H_{N-1}^\dag H_{N-1} \\  \nonumber
 - \lambda_{N-1} (H_{N-1}^\dag H_{N-1})^2 \\ \nonumber
+ \, L_{Fields(N-1)}
+  \,k_{N-1,N} (H_{N-1}^\dag H_{N-1}) (H_{N}^\dag H_N) \\ \nonumber
+  \,(D^{(N)}_{\nu} H_{N})^\dag (D^{(N)\nu}H_{N}) - \mu_{N}^2 H_{N}^\dag H_{N}  - \lambda_{N} (H_{N}^\dag {N})^2  \\ \nonumber 
+ \, L_{Fields(N)}, 
\end{eqnarray}
where $H_1,..., H_{N}$ are the complex scalar fields, $D^{(l)}_{\nu} = \partial_{\nu} - i g^{(l)} V^{(l)}_{\nu}$ is the covariant derivative, $V^{(l)}_{\nu}$ is a gauge field interacting with the scalar field $H_l$,  and $L_{Fields(l)}$ is the Lagrangian of some other fields 
 at the level or link with number $l = 1,\dots , \, N $. 

Suppose all coefficients $\mu_l^2$ except one are equal to zero  $\mu_1^2 = \mu_2^2 = ... = \mu_{N-1}^2 = 0$ and only the last coefficient is nonzero $\mu_{N}^2 \neq 0$ and has a negative value $\mu_{N}^2 =-\mid\mu_{N}^2\mid$. Let also all mixing parameters between the theories at the adjacent levels (links in the chain) be small, namely: 
\begin{eqnarray}
\label{mix-param}
\frac{k_{12}}{\lambda_1}, \,\, 
\frac{k_{12}}{\lambda_2}, \,\,
\frac{k_{23}}{\lambda_2}, \,\,
\frac{k_{23}}{\lambda_3},
\,\,..., \,\, 
\frac{k_{N-1N}}{\lambda_{N-1}} \,\, 
\frac{k_{N-1N}}{\lambda_{N}}\ll \,1.
\end{eqnarray}

The exact formulas for the case of two generations as well as a proof of the diagonalization of the mass matrix $N\times N$ are given in the Appendix.   

\begin{figure*}
\includegraphics[width=0.75\textwidth,clip]
{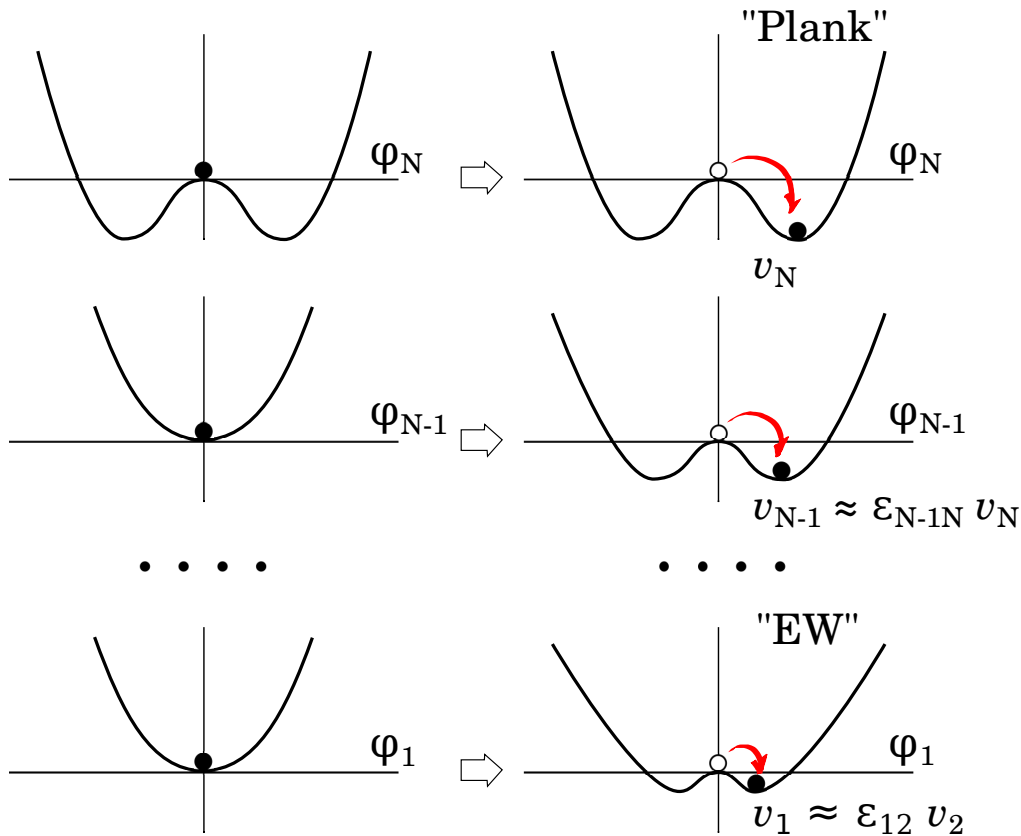}%
\caption{\label{Potential} Illustrative form of potentials before and after spontaneous symmetry breaking induced at the lower levels  by symmetry breaking at the highest level. The lowest level corresponds to the smallest scale, and is designated conventionally as electroweak ("EW"). The uppermost level corresponds to the largest scale, and is conventionally designated as the Planck scale ("Plank"). }
\end{figure*}

The boson $h_N$ at the last step of the chain acquires the mass
\begin{eqnarray}
\label{level_N}
m_N^2 \simeq 2 \lambda_N v_N^2, 
\end{eqnarray}
where 
$v_N^2 = \frac{\mid\mu_{N}^2\mid}{\lambda_N}$ as a result of the usual mechanism of spontaneous symmetry breaking BEH \cite{Higgs:1964pj, Englert:1964et, Guralnik:1964eu}. In this case, due to the mixing, the induced parameter $\widetilde{\mu}_{N-1}^2 =- k_{N-1N} \,\,v_N^2/2 $ appears at the previous step of the chain. With a positive mixing coefficient
$k_{N-1,N}$ at this level spontaneous breaking occurs, the boson $h_{N-1}$ gets mass
\begin{eqnarray}
\label{level_N-1}
m_{N-1}^2 \simeq k_{N-1N}\,\,v_N^2 \simeq \frac{k_{N-1N}}{2\lambda_N}m_N^2,
\end{eqnarray}
and the parameter $\widetilde{\mu}_{N-2}^2$ is induced being proportional to the mixing coefficient $k_{N-2N-1}$ between levels $N-1$ and $N-2$  of the chain. This process continues and, as a result, at the first
step of the chain, the $h_1$-boson mass appears
\begin{eqnarray}
\label{h1-mass}
m_1^2 \simeq \frac{k_{12}}{2\lambda_2} \,\, \frac{k_{23}}{2\lambda_3} \,\,... \,\, \frac{k_{N-1N}}{2\lambda_{N}} \,
m_N^2
\end{eqnarray}
with the induced vacuum expectation value 
\begin{eqnarray}
\label{v1-value}
v_1^2 \simeq \frac{k_{12}}{2\lambda_1} \,\, \frac{k_{23}}{2\lambda_2} \,\,... \,\, \frac{k_{N-1N}}{2\lambda_{N-1}} \, v_N^2
\end{eqnarray}
As one can see, the vacuum expectation value $v_1$ and the Higgs mass $m_1$ can be significantly smaller
than the scale $v_N$ and the mass $m_N$, respectively, due to the product of a number of factors proportional to small mixing coefficients.
This observation could explain the hierarchy between a small scale of the order of $v_1$ and a very large scale of the order of $v_N$. 
The scalar field potentials before and after spontaneous symmetry breaking corresponding to the discussed scenario are illustrated in Fig.~\ref{Potential}.

If we match the smallest vacuum expectation value $v_1$ in the chain with the vacuum expectation value of the Standard Model $v_{SM}$ ($v_1 = v_{SM}$) , the mass $m_1$ of the boson $h_1$ with the SM Higgs boson mass
 $M_{h_{SM}}$ ($m_1 = M_{h_{SM}}$), and the coupling $\lambda_1$
with the SM quartic coupling  $\lambda_{SM}$ ($\lambda_1=\lambda_{SM} $), the SM relation must take place:
\begin{eqnarray}
\label{SM-relation}
M_{h_{SM}}^2 \, = \, 2 \lambda_{SM} v_{SM}^2. 
\end{eqnarray}
It is easy to see that, if we divide the left- and right-hand sides of eq.~(\ref{h1-mass}) by the left- and right-hand sides of 
eq.~(\ref{v1-value}), correspondingly, then taking into account eq.~(\ref{SM-relation}) we get 
$$ 2\lambda_N = \frac{m_N^2}{v_N^2},$$ which exactly matches eq.~(\ref{level_N}), as it should.

Suppose that the largest scale in the chain $v_N$ and the corresponding mass $m_N$ are of the order of Planck scale $\sim 10^{19}$ GeV. Then the product of the factors
\begin{eqnarray}
\label{factors}
\varepsilon_{12} \cdot \varepsilon_{23} \,\,... \,\,\cdot \varepsilon_{i \, i+1} \,\,...\,\, 
\cdot \varepsilon_{N-1 \,N},
 \end{eqnarray}
where $\varepsilon_{i \, i+1} = \sqrt{\frac{k_{i \,i+1}}{2\lambda_{i}}}$, should be about $10^{-17} \div 10^{-16}$ 
assuming that the scale $v_1$ is of the order of the electroweak scale $\sim 10^{2}$ GeV. 

The model (\ref{model_lgrn}) is renormalizable if the Lagrangians  $L_{Fields(i)}$, i=1,...,N involve the operators with dimension 4 or less. On may expect a serious hierarchy problem appearing in the model. Indeed, the correction to the mass squared of the scalar $h_1$ from the loop contribution of the scalar  $h_i$  is proportional to $$m_i^2 \cdot log(m_i^2/m_1^2)$$ leading to a very large shift. However in the model (\ref{model_lgrn}) after the diagonalization the interaction vertex
of the scalars  $h_1$ and $h_i$
contains the coupling constant 
$\varepsilon_{12} \cdot  \,\,... \,\,\cdot \varepsilon_{i-1 \,i}$. Therefore, the loop contribution is proportional to  
$$ \frac{k_{12}}{2\lambda_2} \,\, ... \,\, \frac{k_{i-1i}}{2\lambda_{N}} \, \cdot m_i^2 \cdot log(m_i^2/m_1^2)
\simeq m_1^2 \cdot log(m_i^2/m_1^2), $$
demonstrating that the little hierarchy problem does not show up in the model under consideration. The same argument also holds for the case with possible loop contributions to the scalar $h_1$ mass parameter related to some other fields from $L_{Fields(i)}$ when the potentially large correction  is suppressed by the product of small mixing parameters.

\section{Conclusion}
In this short letter, we discuss a scenario in which a spontaneous symmetry breaking or spontaneous phase transition occurs at some large energy scale. Due to a low mixing, such a phase transition leads to an induced symmetry breaking at some smaller induced scale, which, in turn, due to the next mixing leads to the next induced symmetry breaking at an even smaller scale, and so on. The result is a chain of induced spontaneous symmetry breaking scales. In such a scenario, the large difference between the largest and smallest scales is due to the product of several small factors proportional to the small mixing coefficients, but each of these factors may not be that small. 

In each level of induced symmetry breaking there might be its own gauge theory. Obviously that the interactions of particles at some level with particles at the lowest level will be smaller and smaller 
for higher and higher levels being proportional to smaller and smaller mixing coefficients.
Once the interactions get small enough particles at the corresponding level may play a role of Dark matter particles.

We give an example of a very simple model in which small mixing between the levels in a chain of spontaneous gauge symmetry breaking is due to scalar portals. 
As some speculation, an example is given that the difference between the Planck scale and the electroweak scale can be due to about nine or five induced spontaneous transitions, assuming the same order of small coefficients $10^{-2}$ and $10^{-4}$, respectively. 

Obviously, that having several levels weaker and weaker interacting with the first level associated with the SM, Dark matter might be easily accumulated at far enough levels with small enough couplings to the SM via such a multi-portal scenario.  

We considered an example of a chain with mixing between the nearest neighbors, i.e.  theories at the adjacent levels connected by mixing through scalar portals. Of course, one can think about more complex mixing  not only with the nearest neighbors and not only through scalar portals. In this case, more complex chains with branching and more complex mixing arise.

\section{Acknowledgements}
I am grateful to Sergey Keyzerov and Igor Volobuev for useful discussions.
The study was supported by RFBR and CNRS grant 20-52-15005. The research has been done in the framework  of the Interdisciplinary Scientific and Educational School of Moscow University "Fundamental and Applied Space Research".

\section{Appendix}

Let the chain contains two levels only as given by the eq.~(\ref{2parts_lgrn})
\begin{eqnarray}
\label{2parts_lgrn}
L = (D^{(1)}_{\nu} H_1)^\dag (D^{(1)\nu}H_1) - \mu_1^2 H_1^\dag H_1  - \lambda_1 (H_1^\dag H_1)^2 \,\,\,\,\,\,\,\,\\ \nonumber
+ \,\, k_{12} (H_1^\dag H_1) (H_2^\dag H_2)  \\ \nonumber
+\,\,  (D^{(2)}_{\nu} H_2)^\dag (D^{(2)\nu}H_2) - \mu_2^2 H_2^\dag H_2 - \lambda_2 (H_2^\dag H_2)^2,
\end{eqnarray}
where we set $\mu_1$ equal to zero ($\mu_1^2 = 0$).
In the unitary gauge on both levels the quadratic form of the scalar fields reads as follows:
\begin{eqnarray}
\label{quad_form}
2 \lambda_1 v_1^2 \widetilde{h_1}^2 - 2 k_{12} v_1 v_2 \widetilde{h_1} \widetilde{h_2}
 + 2 \lambda_2 v_2^2 \widetilde{h_2}^2 
\end{eqnarray}
where 
\begin{eqnarray}
\label{v1}
v_1 = \sqrt{\frac{k_{12}}{2 \lambda_1}} v_2, \, \, \, \, \, \,
v_2 = \sqrt{\frac{4 \lambda_1 \mid\mu_{2}^2\mid}{4\lambda_1 \lambda_2 -k_{12}^2 }}
\end{eqnarray}
and the mixing coefficient $k_{12}$ as well as the quartic couplings $\lambda_1$ and  
$\lambda_2$ are chosen to be positive.
Rotation from the unphysical fields $\widetilde{h_1}$ and $\widetilde{h_2}$ to physical fields $h_1$ and $h_2$ with definite masses leads to the following expressions
for the masses squared of eigenstates:
\begin{eqnarray}
\label{masses}
m_{1,2}^2 = \,\,\,\,\,\,\,\,\,\,\,\,\,\,\,\,\,\,\,\,\,\,\,\,\,\,\,\,\,\,\,\,\,\,\,\,\,\,\,\,\,\,\,\,\,\,\,\,\,\,\,\,\,\,\,\,\,\,\,\,\,\,\,\,\,\,\,\,\,\,\,\,\,\,\,\,\,\,\,\,\,\,\,\,\,\,\,\,\,\,\,\,\,\,\\ \nonumber
\left[ \lambda_2 + \frac{k_{12}}{2} 
\mp  \lambda_2 \sqrt{\left(1  - \frac{k_{12}}{2 \lambda_2} \right)^2
 + \frac{k_{12}}{2 \lambda_1} \frac{k_{12}^2}{\lambda_2^2}}\right]  v_2^2.
\end{eqnarray}  
Assuming 
$$\frac{k_{12}}{\lambda_1}, \,\,\frac{k_{12}}{\lambda_2} \,\,\ll \,1$$ 
one gets the following values for the masses:
\begin{eqnarray}
\label{level_12}
m_2^2=2 \lambda_2 v_2^2, \,\, m_1^2 = k_{12}\,\,v_2^2
\end{eqnarray}
in complete agreement with the eqs.~(\ref{level_N}) and (\ref{level_N-1}).

The mixing matrix corresponding to the above example with two levels looks as follows:
\\
\\
\[
\begin{pmatrix}
 2 \lambda_1 v_1^2 & - k_{12} v_1 v_2  \\
- k_{12} v_1 v_2 & 2 \lambda_2 v_2^2  
\end{pmatrix}
\]
\\
This is easily generalized to the model given in (\ref{model_lgrn}). The mixing
matrix has the following form:
\\
\\
\[
\begin{pmatrix}
x_{11} & x_{12} & 0 & 0 & 0 &... & 0    \\
x_{21} &  x_{22} & x_{23} & 0 & 0 &... & 0   \\
0 & x_{32} & x_{33} & x_{34} & 0 & ... & 0  \\
... &... &... &... &... &... & ...\\
0 & 0 & ... & 0 & x_{N-2 N-1} & x_{N-1 N-1} &  x_{N-1 N} \\
0 & 0 & 0 &... & 0 & x_{N N-1} & x_{N N}
\end{pmatrix}
\],
\\
\\
where $x_{11} = 2 \lambda_1 v_1^2$,  $x_{12} = x_{21} = - k_{12} v_1 v_2$, \\
$x_{22} = \lambda_2 v_2^2$, $x_{23}= x_{32} = - k_{23} v_2 v_3$, \\
$x_{33} = 2 \lambda_3 v_3^2$, $x_{34} = x_{43} = - k_{34} v_3 v_4$, \\
$x_{N-1 N-1} = 2 \lambda_{N-1} v_{N-1}^2$,
$x_{N-2 N-1} = x_{N-1 N-2} = - k_{N-2 N-1} v_{N-2} v_{N-1}$,\\
$x_{N N} = 2 \lambda_{N} v_{N}^2$,
$x_{N-1 N} = x_{N N-1} = - k_{N-1 N} v_{N-1} v_{N}$.

Such a N$\times$N matrix can be diagonalized using the product of rotation matrices that sequentially lead to a diagonal form of the 2$\times$2 matrix blocks: 
\\
\\
$
\begin{pmatrix}
cos{\theta_1} & sin{\theta_1}  & 0 & 0 & 0 &... & 0    \\
-sin{\theta_1} & cos{\theta_1} & 0 & 0 &... & 0   \\
0 & 0 & 1 & 0 & 0 & ... & 0  \\
... &... &... &... &... &... & ...\\
0 & 0 & ... & 0 & 0 & 1 &  0 \\
0 & 0 & 0 &... & 0 & 0 & 1
\end{pmatrix}
$
$\times$
\\
$
\begin{pmatrix}
1 & 0  & 0 & 0 & 0 &... & 0    \\
0 & cos{\theta_2} & sin{\theta_2} & 0 & 0 & ... & 0   \\
0 &  -sin{\theta_2} & cos{\theta_2} & 0 & 0 & ... & 0  \\
... &... &... &... &... &... & ...\\
0 & 0 & ... & 0 & 0 & 1 &  0 \\
0 & 0 & 0 &... & 0 & 0 & 1
\end{pmatrix} 
$
$\times$ \\
... \\ \\
$
\begin{pmatrix}
1 & 0  & 0 & 0 & 0 &... & 0    \\
0 & 1 & 0 & 0 & 0 & ... & 0   \\
0 &  0 & 1 & 0 & 0 & ... & 0  \\
... &... &... &... &... &... & ...\\
0 & 0 & ... & 0 & 0 & cos{\theta_N} &   sin{\theta_N} \\
0 & 0 & 0 &... & 0 &  -sin{\theta_N} & cos{\theta_N}
\end{pmatrix}
$, \\
\\
\\
where $\theta_i$, $i=1...N$ are the rotation angles. Under conditions (\ref{mix-param}) and (\ref{level_N}) one gets
expressions  (\ref{level_N-1}) and (\ref{h1-mass}) for the mass eigenstates.    

\FloatBarrier
\bibliography{EWSB_chain_revtex.bib}

\providecommand{\noopsort}[1]{}\providecommand{\singleletter}[1]{#1}%
\begin{thebibliography}{37}%
\makeatletter
\providecommand \@ifxundefined [1]{%
 \@ifx{#1\undefined}
}%
\providecommand \@ifnum [1]{%
 \ifnum #1\expandafter \@firstoftwo
 \else \expandafter \@secondoftwo
 \fi
}%
\providecommand \@ifx [1]{%
 \ifx #1\expandafter \@firstoftwo
 \else \expandafter \@secondoftwo
 \fi
}%
\providecommand \natexlab [1]{#1}%
\providecommand \enquote  [1]{``#1''}%
\providecommand \bibnamefont  [1]{#1}%
\providecommand \bibfnamefont [1]{#1}%
\providecommand \citenamefont [1]{#1}%
\providecommand \href@noop [0]{\@secondoftwo}%
\providecommand \href [0]{\begingroup \@sanitize@url \@href}%
\providecommand \@href[1]{\@@startlink{#1}\@@href}%
\providecommand \@@href[1]{\endgroup#1\@@endlink}%
\providecommand \@sanitize@url [0]{\catcode `\\12\catcode `\$12\catcode
  `\&12\catcode `\#12\catcode `\^12\catcode `\_12\catcode `\%12\relax}%
\providecommand \@@startlink[1]{}%
\providecommand \@@endlink[0]{}%
\providecommand \url  [0]{\begingroup\@sanitize@url \@url }%
\providecommand \@url [1]{\endgroup\@href {#1}{\urlprefix }}%
\providecommand \urlprefix  [0]{URL }%
\providecommand \Eprint [0]{\href }%
\providecommand \doibase [0]{https://doi.org/}%
\providecommand \selectlanguage [0]{\@gobble}%
\providecommand \bibinfo  [0]{\@secondoftwo}%
\providecommand \bibfield  [0]{\@secondoftwo}%
\providecommand \translation [1]{[#1]}%
\providecommand \BibitemOpen [0]{}%
\providecommand \bibitemStop [0]{}%
\providecommand \bibitemNoStop [0]{.\EOS\space}%
\providecommand \EOS [0]{\spacefactor3000\relax}%
\providecommand \BibitemShut  [1]{\csname bibitem#1\endcsname}%
\let\auto@bib@innerbib\@empty
\bibitem [{\citenamefont {Aad}\ \emph {et~al.}(2012)\citenamefont {Aad} \emph
  {et~al.}}]{Aad:2012tfa}%
  \BibitemOpen
  \bibfield  {author} {\bibinfo {author} {\bibfnamefont {G.}~\bibnamefont
  {Aad}} \emph {et~al.} (\bibinfo {collaboration} {ATLAS}),\ }\bibfield
  {title} {\bibinfo {title} {{Observation of a new particle in the search for
  the Standard Model Higgs boson with the ATLAS detector at the LHC}},\ }\href
  {https://doi.org/10.1016/j.physletb.2012.08.020} {\bibfield  {journal}
  {\bibinfo  {journal} {Phys. Lett. B}\ }\textbf {\bibinfo {volume} {716}},\
  \bibinfo {pages} {1} (\bibinfo {year} {2012})},\ \Eprint
  {https://arxiv.org/abs/1207.7214} {arXiv:1207.7214 [hep-ex]} \BibitemShut
  {NoStop}%
\bibitem [{\citenamefont {Chatrchyan}\ \emph {et~al.}(2012)\citenamefont
  {Chatrchyan} \emph {et~al.}}]{Chatrchyan:2012ufa}%
  \BibitemOpen
  \bibfield  {author} {\bibinfo {author} {\bibfnamefont {S.}~\bibnamefont
  {Chatrchyan}} \emph {et~al.} (\bibinfo {collaboration} {CMS}),\ }\bibfield
  {title} {\bibinfo {title} {{Observation of a New Boson at a Mass of 125 GeV
  with the CMS Experiment at the LHC}},\ }\href
  {https://doi.org/10.1016/j.physletb.2012.08.021} {\bibfield  {journal}
  {\bibinfo  {journal} {Phys. Lett. B}\ }\textbf {\bibinfo {volume} {716}},\
  \bibinfo {pages} {30} (\bibinfo {year} {2012})},\ \Eprint
  {https://arxiv.org/abs/1207.7235} {arXiv:1207.7235 [hep-ex]} \BibitemShut
  {NoStop}%
\bibitem [{\citenamefont {Higgs}(1964)}]{Higgs:1964pj}%
  \BibitemOpen
  \bibfield  {author} {\bibinfo {author} {\bibfnamefont {P.~W.}\ \bibnamefont
  {Higgs}},\ }\bibfield  {title} {\bibinfo {title} {{Broken Symmetries and the
  Masses of Gauge Bosons}},\ }\href
  {https://doi.org/10.1103/PhysRevLett.13.508} {\bibfield  {journal} {\bibinfo
  {journal} {Phys. Rev. Lett.}\ }\textbf {\bibinfo {volume} {13}},\ \bibinfo
  {pages} {508} (\bibinfo {year} {1964})}\BibitemShut {NoStop}%
\bibitem [{\citenamefont {Englert}\ and\ \citenamefont
  {Brout}(1964)}]{Englert:1964et}%
  \BibitemOpen
  \bibfield  {author} {\bibinfo {author} {\bibfnamefont {F.}~\bibnamefont
  {Englert}}\ and\ \bibinfo {author} {\bibfnamefont {R.}~\bibnamefont
  {Brout}},\ }\bibfield  {title} {\bibinfo {title} {{Broken Symmetry and the
  Mass of Gauge Vector Mesons}},\ }\href
  {https://doi.org/10.1103/PhysRevLett.13.321} {\bibfield  {journal} {\bibinfo
  {journal} {Phys. Rev. Lett.}\ }\textbf {\bibinfo {volume} {13}},\ \bibinfo
  {pages} {321} (\bibinfo {year} {1964})}\BibitemShut {NoStop}%
\bibitem [{\citenamefont {Guralnik}\ \emph {et~al.}(1964)\citenamefont
  {Guralnik}, \citenamefont {Hagen},\ and\ \citenamefont
  {Kibble}}]{Guralnik:1964eu}%
  \BibitemOpen
  \bibfield  {author} {\bibinfo {author} {\bibfnamefont {G.~S.}\ \bibnamefont
  {Guralnik}}, \bibinfo {author} {\bibfnamefont {C.~R.}\ \bibnamefont
  {Hagen}},\ and\ \bibinfo {author} {\bibfnamefont {T.~W.~B.}\ \bibnamefont
  {Kibble}},\ }\bibfield  {title} {\bibinfo {title} {{Global Conservation Laws
  and Massless Particles}},\ }\href
  {https://doi.org/10.1103/PhysRevLett.13.585} {\bibfield  {journal} {\bibinfo
  {journal} {Phys. Rev. Lett.}\ }\textbf {\bibinfo {volume} {13}},\ \bibinfo
  {pages} {585} (\bibinfo {year} {1964})}\BibitemShut {NoStop}%
\bibitem [{\citenamefont {Lee}\ \emph {et~al.}(1977{\natexlab{a}})\citenamefont
  {Lee}, \citenamefont {Quigg},\ and\ \citenamefont {Thacker}}]{Lee:1977yc}%
  \BibitemOpen
  \bibfield  {author} {\bibinfo {author} {\bibfnamefont {B.~W.}\ \bibnamefont
  {Lee}}, \bibinfo {author} {\bibfnamefont {C.}~\bibnamefont {Quigg}},\ and\
  \bibinfo {author} {\bibfnamefont {H.~B.}\ \bibnamefont {Thacker}},\
  }\bibfield  {title} {\bibinfo {title} {{The Strength of Weak Interactions at
  Very High-Energies and the Higgs Boson Mass}},\ }\href
  {https://doi.org/10.1103/PhysRevLett.38.883} {\bibfield  {journal} {\bibinfo
  {journal} {Phys. Rev. Lett.}\ }\textbf {\bibinfo {volume} {38}},\ \bibinfo
  {pages} {883} (\bibinfo {year} {1977}{\natexlab{a}})}\BibitemShut {NoStop}%
\bibitem [{\citenamefont {Lee}\ \emph {et~al.}(1977{\natexlab{b}})\citenamefont
  {Lee}, \citenamefont {Quigg},\ and\ \citenamefont {Thacker}}]{Lee:1977eg}%
  \BibitemOpen
  \bibfield  {author} {\bibinfo {author} {\bibfnamefont {B.~W.}\ \bibnamefont
  {Lee}}, \bibinfo {author} {\bibfnamefont {C.}~\bibnamefont {Quigg}},\ and\
  \bibinfo {author} {\bibfnamefont {H.~B.}\ \bibnamefont {Thacker}},\
  }\bibfield  {title} {\bibinfo {title} {{Weak Interactions at Very
  High-Energies: The Role of the Higgs Boson Mass}},\ }\href
  {https://doi.org/10.1103/PhysRevD.16.1519} {\bibfield  {journal} {\bibinfo
  {journal} {Phys. Rev. D}\ }\textbf {\bibinfo {volume} {16}},\ \bibinfo
  {pages} {1519} (\bibinfo {year} {1977}{\natexlab{b}})}\BibitemShut {NoStop}%
\bibitem [{\citenamefont {Chanowitz}(1986)}]{Chanowitz:1986ar}%
  \BibitemOpen
  \bibfield  {author} {\bibinfo {author} {\bibfnamefont {M.~S.}\ \bibnamefont
  {Chanowitz}},\ }\bibfield  {title} {\bibinfo {title} {{Universal $W$, $Z$
  Scattering Theorems and No Lose Corollary for the {SSC}}},\ }in\ \href@noop
  {} {\emph {\bibinfo {booktitle} {{23rd International Conference on
  High-Energy Physics}}}}\ (\bibinfo {year} {1986})\BibitemShut {NoStop}%
\bibitem [{\citenamefont {Dicus}\ and\ \citenamefont
  {Mathur}(1973)}]{Dicus:1992vj}%
  \BibitemOpen
  \bibfield  {author} {\bibinfo {author} {\bibfnamefont {D.~A.}\ \bibnamefont
  {Dicus}}\ and\ \bibinfo {author} {\bibfnamefont {V.~S.}\ \bibnamefont
  {Mathur}},\ }\bibfield  {title} {\bibinfo {title} {{Upper bounds on the
  values of masses in unified gauge theories}},\ }\href
  {https://doi.org/10.1103/PhysRevD.7.3111} {\bibfield  {journal} {\bibinfo
  {journal} {Phys. Rev. D}\ }\textbf {\bibinfo {volume} {7}},\ \bibinfo {pages}
  {3111} (\bibinfo {year} {1973})}\BibitemShut {NoStop}%
\bibitem [{\citenamefont {Silveira}\ and\ \citenamefont
  {Zee}(1985)}]{Silveira:1985rk}%
  \BibitemOpen
  \bibfield  {author} {\bibinfo {author} {\bibfnamefont {V.}~\bibnamefont
  {Silveira}}\ and\ \bibinfo {author} {\bibfnamefont {A.}~\bibnamefont {Zee}},\
  }\bibfield  {title} {\bibinfo {title} {{SCALAR PHANTOMS}},\ }\href
  {https://doi.org/10.1016/0370-2693(85)90624-0} {\bibfield  {journal}
  {\bibinfo  {journal} {Phys. Lett. B}\ }\textbf {\bibinfo {volume} {161}},\
  \bibinfo {pages} {136} (\bibinfo {year} {1985})}\BibitemShut {NoStop}%
\bibitem [{\citenamefont {McDonald}(1994)}]{McDonald:1993ex}%
  \BibitemOpen
  \bibfield  {author} {\bibinfo {author} {\bibfnamefont {J.}~\bibnamefont
  {McDonald}},\ }\bibfield  {title} {\bibinfo {title} {{Gauge singlet scalars
  as cold dark matter}},\ }\href {https://doi.org/10.1103/PhysRevD.50.3637}
  {\bibfield  {journal} {\bibinfo  {journal} {Phys. Rev. D}\ }\textbf {\bibinfo
  {volume} {50}},\ \bibinfo {pages} {3637} (\bibinfo {year} {1994})},\ \Eprint
  {https://arxiv.org/abs/hep-ph/0702143} {arXiv:hep-ph/0702143} \BibitemShut
  {NoStop}%
\bibitem [{\citenamefont {Burgess}\ \emph {et~al.}(2001)\citenamefont
  {Burgess}, \citenamefont {Pospelov},\ and\ \citenamefont {ter
  Veldhuis}}]{Burgess:2000yq}%
  \BibitemOpen
  \bibfield  {author} {\bibinfo {author} {\bibfnamefont {C.~P.}\ \bibnamefont
  {Burgess}}, \bibinfo {author} {\bibfnamefont {M.}~\bibnamefont {Pospelov}},\
  and\ \bibinfo {author} {\bibfnamefont {T.}~\bibnamefont {ter Veldhuis}},\
  }\bibfield  {title} {\bibinfo {title} {{The Minimal model of nonbaryonic dark
  matter: A Singlet scalar}},\ }\href
  {https://doi.org/10.1016/S0550-3213(01)00513-2} {\bibfield  {journal}
  {\bibinfo  {journal} {Nucl. Phys. B}\ }\textbf {\bibinfo {volume} {619}},\
  \bibinfo {pages} {709} (\bibinfo {year} {2001})},\ \Eprint
  {https://arxiv.org/abs/hep-ph/0011335} {arXiv:hep-ph/0011335} \BibitemShut
  {NoStop}%
\bibitem [{\citenamefont {Davoudiasl}\ \emph {et~al.}(2005)\citenamefont
  {Davoudiasl}, \citenamefont {Kitano}, \citenamefont {Li},\ and\ \citenamefont
  {Murayama}}]{Davoudiasl:2004be}%
  \BibitemOpen
  \bibfield  {author} {\bibinfo {author} {\bibfnamefont {H.}~\bibnamefont
  {Davoudiasl}}, \bibinfo {author} {\bibfnamefont {R.}~\bibnamefont {Kitano}},
  \bibinfo {author} {\bibfnamefont {T.}~\bibnamefont {Li}},\ and\ \bibinfo
  {author} {\bibfnamefont {H.}~\bibnamefont {Murayama}},\ }\bibfield  {title}
  {\bibinfo {title} {{The New minimal standard model}},\ }\href
  {https://doi.org/10.1016/j.physletb.2005.01.026} {\bibfield  {journal}
  {\bibinfo  {journal} {Phys. Lett. B}\ }\textbf {\bibinfo {volume} {609}},\
  \bibinfo {pages} {117} (\bibinfo {year} {2005})},\ \Eprint
  {https://arxiv.org/abs/hep-ph/0405097} {arXiv:hep-ph/0405097} \BibitemShut
  {NoStop}%
\bibitem [{\citenamefont {Schabinger}\ and\ \citenamefont
  {Wells}(2005)}]{Schabinger:2005ei}%
  \BibitemOpen
  \bibfield  {author} {\bibinfo {author} {\bibfnamefont {R.~M.}\ \bibnamefont
  {Schabinger}}\ and\ \bibinfo {author} {\bibfnamefont {J.~D.}\ \bibnamefont
  {Wells}},\ }\bibfield  {title} {\bibinfo {title} {{A Minimal spontaneously
  broken hidden sector and its impact on Higgs boson physics at the large
  hadron collider}},\ }\href {https://doi.org/10.1103/PhysRevD.72.093007}
  {\bibfield  {journal} {\bibinfo  {journal} {Phys. Rev. D}\ }\textbf {\bibinfo
  {volume} {72}},\ \bibinfo {pages} {093007} (\bibinfo {year} {2005})},\
  \Eprint {https://arxiv.org/abs/hep-ph/0509209} {arXiv:hep-ph/0509209}
  \BibitemShut {NoStop}%
\bibitem [{\citenamefont {Barbieri}\ \emph {et~al.}(2005)\citenamefont
  {Barbieri}, \citenamefont {Gregoire},\ and\ \citenamefont
  {Hall}}]{Barbieri:2005ri}%
  \BibitemOpen
  \bibfield  {author} {\bibinfo {author} {\bibfnamefont {R.}~\bibnamefont
  {Barbieri}}, \bibinfo {author} {\bibfnamefont {T.}~\bibnamefont {Gregoire}},\
  and\ \bibinfo {author} {\bibfnamefont {L.~J.}\ \bibnamefont {Hall}},\
  }\bibfield  {title} {\bibinfo {title} {{\rm Mirror world at the large hadron
  collider}},\ }\Eprint {https://arxiv.org/abs/hep-ph/0509242}
  {arXiv:hep-ph/0509242}  (\bibinfo {year} {2005})\BibitemShut {NoStop}%
\bibitem [{\citenamefont {Patt}\ and\ \citenamefont
  {Wilczek}(2006)}]{Patt:2006fw}%
  \BibitemOpen
  \bibfield  {author} {\bibinfo {author} {\bibfnamefont {B.}~\bibnamefont
  {Patt}}\ and\ \bibinfo {author} {\bibfnamefont {F.}~\bibnamefont {Wilczek}},\
  }\bibfield  {title} {\bibinfo {title} {{\rm Higgs-field portal into hidden
  sectors}},\ }\Eprint {https://arxiv.org/abs/hep-ph/0605188}
  {arXiv:hep-ph/0605188}  (\bibinfo {year} {2006})\BibitemShut {NoStop}%
\bibitem [{\citenamefont {Strassler}\ and\ \citenamefont
  {Zurek}(2008)}]{Strassler:2006ri}%
  \BibitemOpen
  \bibfield  {author} {\bibinfo {author} {\bibfnamefont {M.~J.}\ \bibnamefont
  {Strassler}}\ and\ \bibinfo {author} {\bibfnamefont {K.~M.}\ \bibnamefont
  {Zurek}},\ }\bibfield  {title} {\bibinfo {title} {{Discovering the Higgs
  through highly-displaced vertices}},\ }\href
  {https://doi.org/10.1016/j.physletb.2008.02.008} {\bibfield  {journal}
  {\bibinfo  {journal} {Phys. Lett. B}\ }\textbf {\bibinfo {volume} {661}},\
  \bibinfo {pages} {263} (\bibinfo {year} {2008})},\ \Eprint
  {https://arxiv.org/abs/hep-ph/0605193} {arXiv:hep-ph/0605193} \BibitemShut
  {NoStop}%
\bibitem [{\citenamefont {Bowen}\ \emph {et~al.}(2007)\citenamefont {Bowen},
  \citenamefont {Cui},\ and\ \citenamefont {Wells}}]{Bowen:2007ia}%
  \BibitemOpen
  \bibfield  {author} {\bibinfo {author} {\bibfnamefont {M.}~\bibnamefont
  {Bowen}}, \bibinfo {author} {\bibfnamefont {Y.}~\bibnamefont {Cui}},\ and\
  \bibinfo {author} {\bibfnamefont {J.~D.}\ \bibnamefont {Wells}},\ }\bibfield
  {title} {\bibinfo {title} {{Narrow trans-TeV Higgs bosons and H
  ---\ensuremath{>} hh decays: Two LHC search paths for a hidden sector Higgs
  boson}},\ }\href {https://doi.org/10.1088/1126-6708/2007/03/036} {\bibfield
  {journal} {\bibinfo  {journal} {JHEP}\ }\textbf {\bibinfo {volume} {03}},\
  \bibinfo {pages} {036}},\ \Eprint {https://arxiv.org/abs/hep-ph/0701035}
  {arXiv:hep-ph/0701035} \BibitemShut {NoStop}%
\bibitem [{\citenamefont {Barger}\ \emph {et~al.}(2008)\citenamefont {Barger},
  \citenamefont {Langacker}, \citenamefont {McCaskey}, \citenamefont
  {Ramsey-Musolf},\ and\ \citenamefont {Shaughnessy}}]{Barger:2007im}%
  \BibitemOpen
  \bibfield  {author} {\bibinfo {author} {\bibfnamefont {V.}~\bibnamefont
  {Barger}}, \bibinfo {author} {\bibfnamefont {P.}~\bibnamefont {Langacker}},
  \bibinfo {author} {\bibfnamefont {M.}~\bibnamefont {McCaskey}}, \bibinfo
  {author} {\bibfnamefont {M.~J.}\ \bibnamefont {Ramsey-Musolf}},\ and\
  \bibinfo {author} {\bibfnamefont {G.}~\bibnamefont {Shaughnessy}},\
  }\bibfield  {title} {\bibinfo {title} {{LHC Phenomenology of an Extended
  Standard Model with a Real Scalar Singlet}},\ }\href
  {https://doi.org/10.1103/PhysRevD.77.035005} {\bibfield  {journal} {\bibinfo
  {journal} {Phys. Rev. D}\ }\textbf {\bibinfo {volume} {77}},\ \bibinfo
  {pages} {035005} (\bibinfo {year} {2008})},\ \Eprint
  {https://arxiv.org/abs/0706.4311} {arXiv:0706.4311 [hep-ph]} \BibitemShut
  {NoStop}%
\bibitem [{\citenamefont {Barger}\ \emph {et~al.}(2009)\citenamefont {Barger},
  \citenamefont {Langacker}, \citenamefont {McCaskey}, \citenamefont
  {Ramsey-Musolf},\ and\ \citenamefont {Shaughnessy}}]{Barger:2008jx}%
  \BibitemOpen
  \bibfield  {author} {\bibinfo {author} {\bibfnamefont {V.}~\bibnamefont
  {Barger}}, \bibinfo {author} {\bibfnamefont {P.}~\bibnamefont {Langacker}},
  \bibinfo {author} {\bibfnamefont {M.}~\bibnamefont {McCaskey}}, \bibinfo
  {author} {\bibfnamefont {M.}~\bibnamefont {Ramsey-Musolf}},\ and\ \bibinfo
  {author} {\bibfnamefont {G.}~\bibnamefont {Shaughnessy}},\ }\bibfield
  {title} {\bibinfo {title} {{Complex Singlet Extension of the Standard
  Model}},\ }\href {https://doi.org/10.1103/PhysRevD.79.015018} {\bibfield
  {journal} {\bibinfo  {journal} {Phys. Rev. D}\ }\textbf {\bibinfo {volume}
  {79}},\ \bibinfo {pages} {015018} (\bibinfo {year} {2009})},\ \Eprint
  {https://arxiv.org/abs/0811.0393} {arXiv:0811.0393 [hep-ph]} \BibitemShut
  {NoStop}%
\bibitem [{\citenamefont {He}\ \emph {et~al.}(2009)\citenamefont {He},
  \citenamefont {Li}, \citenamefont {Li}, \citenamefont {Tandean},\ and\
  \citenamefont {Tsai}}]{He:2008qm}%
  \BibitemOpen
  \bibfield  {author} {\bibinfo {author} {\bibfnamefont {X.-G.}\ \bibnamefont
  {He}}, \bibinfo {author} {\bibfnamefont {T.}~\bibnamefont {Li}}, \bibinfo
  {author} {\bibfnamefont {X.-Q.}\ \bibnamefont {Li}}, \bibinfo {author}
  {\bibfnamefont {J.}~\bibnamefont {Tandean}},\ and\ \bibinfo {author}
  {\bibfnamefont {H.-C.}\ \bibnamefont {Tsai}},\ }\bibfield  {title} {\bibinfo
  {title} {{Constraints on Scalar Dark Matter from Direct Experimental
  Searches}},\ }\href {https://doi.org/10.1103/PhysRevD.79.023521} {\bibfield
  {journal} {\bibinfo  {journal} {Phys. Rev. D}\ }\textbf {\bibinfo {volume}
  {79}},\ \bibinfo {pages} {023521} (\bibinfo {year} {2009})},\ \Eprint
  {https://arxiv.org/abs/0811.0658} {arXiv:0811.0658 [hep-ph]} \BibitemShut
  {NoStop}%
\bibitem [{\citenamefont {Bock}\ \emph {et~al.}(2011)\citenamefont {Bock},
  \citenamefont {Lafaye}, \citenamefont {Plehn}, \citenamefont {Rauch},
  \citenamefont {Zerwas},\ and\ \citenamefont {Zerwas}}]{Bock:2010nz}%
  \BibitemOpen
  \bibfield  {author} {\bibinfo {author} {\bibfnamefont {S.}~\bibnamefont
  {Bock}}, \bibinfo {author} {\bibfnamefont {R.}~\bibnamefont {Lafaye}},
  \bibinfo {author} {\bibfnamefont {T.}~\bibnamefont {Plehn}}, \bibinfo
  {author} {\bibfnamefont {M.}~\bibnamefont {Rauch}}, \bibinfo {author}
  {\bibfnamefont {D.}~\bibnamefont {Zerwas}},\ and\ \bibinfo {author}
  {\bibfnamefont {P.~M.}\ \bibnamefont {Zerwas}},\ }\bibfield  {title}
  {\bibinfo {title} {{Measuring Hidden Higgs and Strongly-Interacting Higgs
  Scenarios}},\ }\href {https://doi.org/10.1016/j.physletb.2010.09.032}
  {\bibfield  {journal} {\bibinfo  {journal} {Phys. Lett. B}\ }\textbf
  {\bibinfo {volume} {694}},\ \bibinfo {pages} {44} (\bibinfo {year} {2011})},\
  \Eprint {https://arxiv.org/abs/1007.2645} {arXiv:1007.2645 [hep-ph]}
  \BibitemShut {NoStop}%
\bibitem [{\citenamefont {Englert}\ \emph {et~al.}(2011)\citenamefont
  {Englert}, \citenamefont {Plehn}, \citenamefont {Zerwas},\ and\ \citenamefont
  {Zerwas}}]{Englert:2011yb}%
  \BibitemOpen
  \bibfield  {author} {\bibinfo {author} {\bibfnamefont {C.}~\bibnamefont
  {Englert}}, \bibinfo {author} {\bibfnamefont {T.}~\bibnamefont {Plehn}},
  \bibinfo {author} {\bibfnamefont {D.}~\bibnamefont {Zerwas}},\ and\ \bibinfo
  {author} {\bibfnamefont {P.~M.}\ \bibnamefont {Zerwas}},\ }\bibfield  {title}
  {\bibinfo {title} {{Exploring the Higgs portal}},\ }\href
  {https://doi.org/10.1016/j.physletb.2011.08.002} {\bibfield  {journal}
  {\bibinfo  {journal} {Phys. Lett. B}\ }\textbf {\bibinfo {volume} {703}},\
  \bibinfo {pages} {298} (\bibinfo {year} {2011})},\ \Eprint
  {https://arxiv.org/abs/1106.3097} {arXiv:1106.3097 [hep-ph]} \BibitemShut
  {NoStop}%
\bibitem [{\citenamefont {Lebedev}\ and\ \citenamefont
  {Lee}(2011)}]{Lebedev:2011aq}%
  \BibitemOpen
  \bibfield  {author} {\bibinfo {author} {\bibfnamefont {O.}~\bibnamefont
  {Lebedev}}\ and\ \bibinfo {author} {\bibfnamefont {H.~M.}\ \bibnamefont
  {Lee}},\ }\bibfield  {title} {\bibinfo {title} {{Higgs Portal Inflation}},\
  }\href {https://doi.org/10.1140/epjc/s10052-011-1821-0} {\bibfield  {journal}
  {\bibinfo  {journal} {Eur. Phys. J. C}\ }\textbf {\bibinfo {volume} {71}},\
  \bibinfo {pages} {1821} (\bibinfo {year} {2011})},\ \Eprint
  {https://arxiv.org/abs/1105.2284} {arXiv:1105.2284 [hep-ph]} \BibitemShut
  {NoStop}%
\bibitem [{\citenamefont {Mambrini}(2011)}]{Mambrini:2011ik}%
  \BibitemOpen
  \bibfield  {author} {\bibinfo {author} {\bibfnamefont {Y.}~\bibnamefont
  {Mambrini}},\ }\bibfield  {title} {\bibinfo {title} {{Higgs searches and
  singlet scalar dark matter: Combined constraints from XENON 100 and the
  LHC}},\ }\href {https://doi.org/10.1103/PhysRevD.84.115017} {\bibfield
  {journal} {\bibinfo  {journal} {Phys. Rev. D}\ }\textbf {\bibinfo {volume}
  {84}},\ \bibinfo {pages} {115017} (\bibinfo {year} {2011})},\ \Eprint
  {https://arxiv.org/abs/1108.0671} {arXiv:1108.0671 [hep-ph]} \BibitemShut
  {NoStop}%
\bibitem [{\citenamefont {Belanger}\ \emph {et~al.}(2013)\citenamefont
  {Belanger}, \citenamefont {Kannike}, \citenamefont {Pukhov},\ and\
  \citenamefont {Raidal}}]{Belanger:2012zr}%
  \BibitemOpen
  \bibfield  {author} {\bibinfo {author} {\bibfnamefont {G.}~\bibnamefont
  {Belanger}}, \bibinfo {author} {\bibfnamefont {K.}~\bibnamefont {Kannike}},
  \bibinfo {author} {\bibfnamefont {A.}~\bibnamefont {Pukhov}},\ and\ \bibinfo
  {author} {\bibfnamefont {M.}~\bibnamefont {Raidal}},\ }\bibfield  {title}
  {\bibinfo {title} {{$Z_3$ Scalar Singlet Dark Matter}},\ }\href
  {https://doi.org/10.1088/1475-7516/2013/01/022} {\bibfield  {journal}
  {\bibinfo  {journal} {JCAP}\ }\textbf {\bibinfo {volume} {01}},\ \bibinfo
  {pages} {022}},\ \Eprint {https://arxiv.org/abs/1211.1014} {arXiv:1211.1014
  [hep-ph]} \BibitemShut {NoStop}%
\bibitem [{\citenamefont {Cline}\ \emph {et~al.}(2013)\citenamefont {Cline},
  \citenamefont {Kainulainen}, \citenamefont {Scott},\ and\ \citenamefont
  {Weniger}}]{Cline:2013gha}%
  \BibitemOpen
  \bibfield  {author} {\bibinfo {author} {\bibfnamefont {J.~M.}\ \bibnamefont
  {Cline}}, \bibinfo {author} {\bibfnamefont {K.}~\bibnamefont {Kainulainen}},
  \bibinfo {author} {\bibfnamefont {P.}~\bibnamefont {Scott}},\ and\ \bibinfo
  {author} {\bibfnamefont {C.}~\bibnamefont {Weniger}},\ }\bibfield  {title}
  {\bibinfo {title} {{Update on scalar singlet dark matter}},\ }\href
  {https://doi.org/10.1103/PhysRevD.88.055025} {\bibfield  {journal} {\bibinfo
  {journal} {Phys. Rev. D}\ }\textbf {\bibinfo {volume} {88}},\ \bibinfo
  {pages} {055025} (\bibinfo {year} {2013})},\ \bibinfo {note} {[Erratum:
  Phys.Rev.D 92, 039906 (2015)]},\ \Eprint {https://arxiv.org/abs/1306.4710}
  {arXiv:1306.4710 [hep-ph]} \BibitemShut {NoStop}%
\bibitem [{\citenamefont {Gabrielli}\ \emph {et~al.}(2014)\citenamefont
  {Gabrielli}, \citenamefont {Heikinheimo}, \citenamefont {Kannike},
  \citenamefont {Racioppi}, \citenamefont {Raidal},\ and\ \citenamefont
  {Spethmann}}]{Gabrielli:2013hma}%
  \BibitemOpen
  \bibfield  {author} {\bibinfo {author} {\bibfnamefont {E.}~\bibnamefont
  {Gabrielli}}, \bibinfo {author} {\bibfnamefont {M.}~\bibnamefont
  {Heikinheimo}}, \bibinfo {author} {\bibfnamefont {K.}~\bibnamefont
  {Kannike}}, \bibinfo {author} {\bibfnamefont {A.}~\bibnamefont {Racioppi}},
  \bibinfo {author} {\bibfnamefont {M.}~\bibnamefont {Raidal}},\ and\ \bibinfo
  {author} {\bibfnamefont {C.}~\bibnamefont {Spethmann}},\ }\bibfield  {title}
  {\bibinfo {title} {{Towards Completing the Standard Model: Vacuum Stability,
  EWSB and Dark Matter}},\ }\href {https://doi.org/10.1103/PhysRevD.89.015017}
  {\bibfield  {journal} {\bibinfo  {journal} {Phys. Rev. D}\ }\textbf {\bibinfo
  {volume} {89}},\ \bibinfo {pages} {015017} (\bibinfo {year} {2014})},\
  \Eprint {https://arxiv.org/abs/1309.6632} {arXiv:1309.6632 [hep-ph]}
  \BibitemShut {NoStop}%
\bibitem [{\citenamefont {Falkowski}\ \emph {et~al.}(2015)\citenamefont
  {Falkowski}, \citenamefont {Gross},\ and\ \citenamefont
  {Lebedev}}]{Falkowski:2015iwa}%
  \BibitemOpen
  \bibfield  {author} {\bibinfo {author} {\bibfnamefont {A.}~\bibnamefont
  {Falkowski}}, \bibinfo {author} {\bibfnamefont {C.}~\bibnamefont {Gross}},\
  and\ \bibinfo {author} {\bibfnamefont {O.}~\bibnamefont {Lebedev}},\
  }\bibfield  {title} {\bibinfo {title} {{A second Higgs from the Higgs
  portal}},\ }\href {https://doi.org/10.1007/JHEP05(2015)057} {\bibfield
  {journal} {\bibinfo  {journal} {JHEP}\ }\textbf {\bibinfo {volume} {05}},\
  \bibinfo {pages} {057}},\ \Eprint {https://arxiv.org/abs/1502.01361}
  {arXiv:1502.01361 [hep-ph]} \BibitemShut {NoStop}%
\bibitem [{\citenamefont {Robens}\ and\ \citenamefont
  {Stefaniak}(2015)}]{Robens:2015gla}%
  \BibitemOpen
  \bibfield  {author} {\bibinfo {author} {\bibfnamefont {T.}~\bibnamefont
  {Robens}}\ and\ \bibinfo {author} {\bibfnamefont {T.}~\bibnamefont
  {Stefaniak}},\ }\bibfield  {title} {\bibinfo {title} {{Status of the Higgs
  Singlet Extension of the Standard Model after LHC Run 1}},\ }\href
  {https://doi.org/10.1140/epjc/s10052-015-3323-y} {\bibfield  {journal}
  {\bibinfo  {journal} {Eur. Phys. J. C}\ }\textbf {\bibinfo {volume} {75}},\
  \bibinfo {pages} {104} (\bibinfo {year} {2015})},\ \Eprint
  {https://arxiv.org/abs/1501.02234} {arXiv:1501.02234 [hep-ph]} \BibitemShut
  {NoStop}%
\bibitem [{\citenamefont {Mart\'\i{}n~Lozano}\ \emph
  {et~al.}(2015)\citenamefont {Mart\'\i{}n~Lozano}, \citenamefont {Moreno},\
  and\ \citenamefont {Park}}]{Martin-Lozano:2015dja}%
  \BibitemOpen
  \bibfield  {author} {\bibinfo {author} {\bibfnamefont {V.}~\bibnamefont
  {Mart\'\i{}n~Lozano}}, \bibinfo {author} {\bibfnamefont {J.~M.}\ \bibnamefont
  {Moreno}},\ and\ \bibinfo {author} {\bibfnamefont {C.~B.}\ \bibnamefont
  {Park}},\ }\bibfield  {title} {\bibinfo {title} {{Resonant Higgs boson pair
  production in the $ hh\to b\overline{b}\ WW\to b\overline{b}{\ell}^{+}\nu
  {\ell}^{-}\overline{\nu} $ decay channel}},\ }\href
  {https://doi.org/10.1007/JHEP08(2015)004} {\bibfield  {journal} {\bibinfo
  {journal} {JHEP}\ }\textbf {\bibinfo {volume} {08}},\ \bibinfo {pages}
  {004}},\ \Eprint {https://arxiv.org/abs/1501.03799} {arXiv:1501.03799
  [hep-ph]} \BibitemShut {NoStop}%
\bibitem [{\citenamefont {Casas}\ \emph {et~al.}(2017)\citenamefont {Casas},
  \citenamefont {Cerde\~no}, \citenamefont {Moreno},\ and\ \citenamefont
  {Quilis}}]{Casas:2017jjg}%
  \BibitemOpen
  \bibfield  {author} {\bibinfo {author} {\bibfnamefont {J.~A.}\ \bibnamefont
  {Casas}}, \bibinfo {author} {\bibfnamefont {D.~G.}\ \bibnamefont
  {Cerde\~no}}, \bibinfo {author} {\bibfnamefont {J.~M.}\ \bibnamefont
  {Moreno}},\ and\ \bibinfo {author} {\bibfnamefont {J.}~\bibnamefont
  {Quilis}},\ }\bibfield  {title} {\bibinfo {title} {{Reopening the Higgs
  portal for single scalar dark matter}},\ }\href
  {https://doi.org/10.1007/JHEP05(2017)036} {\bibfield  {journal} {\bibinfo
  {journal} {JHEP}\ }\textbf {\bibinfo {volume} {05}},\ \bibinfo {pages}
  {036}},\ \Eprint {https://arxiv.org/abs/1701.08134} {arXiv:1701.08134
  [hep-ph]} \BibitemShut {NoStop}%
\bibitem [{\citenamefont {Athron}\ \emph {et~al.}(2017)\citenamefont {Athron}
  \emph {et~al.}}]{Athron:2017kgt}%
  \BibitemOpen
  \bibfield  {author} {\bibinfo {author} {\bibfnamefont {P.}~\bibnamefont
  {Athron}} \emph {et~al.} (\bibinfo {collaboration} {GAMBIT}),\ }\bibfield
  {title} {\bibinfo {title} {{Status of the scalar singlet dark matter
  model}},\ }\href {https://doi.org/10.1140/epjc/s10052-017-5113-1} {\bibfield
  {journal} {\bibinfo  {journal} {Eur. Phys. J. C}\ }\textbf {\bibinfo {volume}
  {77}},\ \bibinfo {pages} {568} (\bibinfo {year} {2017})},\ \Eprint
  {https://arxiv.org/abs/1705.07931} {arXiv:1705.07931 [hep-ph]} \BibitemShut
  {NoStop}%
\bibitem [{\citenamefont {Arcadi}\ \emph {et~al.}(2020)\citenamefont {Arcadi},
  \citenamefont {Djouadi},\ and\ \citenamefont {Raidal}}]{Arcadi:2019lka}%
  \BibitemOpen
  \bibfield  {author} {\bibinfo {author} {\bibfnamefont {G.}~\bibnamefont
  {Arcadi}}, \bibinfo {author} {\bibfnamefont {A.}~\bibnamefont {Djouadi}},\
  and\ \bibinfo {author} {\bibfnamefont {M.}~\bibnamefont {Raidal}},\
  }\bibfield  {title} {\bibinfo {title} {{Dark Matter through the Higgs
  portal}},\ }\href {https://doi.org/10.1016/j.physrep.2019.11.003} {\bibfield
  {journal} {\bibinfo  {journal} {Phys. Rept.}\ }\textbf {\bibinfo {volume}
  {842}},\ \bibinfo {pages} {1} (\bibinfo {year} {2020})},\ \Eprint
  {https://arxiv.org/abs/1903.03616} {arXiv:1903.03616 [hep-ph]} \BibitemShut
  {NoStop}%
\bibitem [{\citenamefont {Kannike}\ \emph {et~al.}(2021)\citenamefont
  {Kannike}, \citenamefont {Koivunen},\ and\ \citenamefont
  {Raidal}}]{Kannike:2020qtw}%
  \BibitemOpen
  \bibfield  {author} {\bibinfo {author} {\bibfnamefont {K.}~\bibnamefont
  {Kannike}}, \bibinfo {author} {\bibfnamefont {N.}~\bibnamefont {Koivunen}},\
  and\ \bibinfo {author} {\bibfnamefont {M.}~\bibnamefont {Raidal}},\
  }\bibfield  {title} {\bibinfo {title} {{Principle of Multiple Point
  Criticality in Multi-Scalar Dark Matter Models}},\ }\href
  {https://doi.org/10.1016/j.nuclphysb.2021.115441} {\bibfield  {journal}
  {\bibinfo  {journal} {Nucl. Phys. B}\ }\textbf {\bibinfo {volume} {968}},\
  \bibinfo {pages} {115441} (\bibinfo {year} {2021})},\ \Eprint
  {https://arxiv.org/abs/2010.09718} {arXiv:2010.09718 [hep-ph]} \BibitemShut
  {NoStop}%
\bibitem [{\citenamefont {D\'\i{}az~S\'aez}\ \emph {et~al.}(2021)\citenamefont
  {D\'\i{}az~S\'aez}, \citenamefont {M\"ohling},\ and\ \citenamefont
  {St\"ockinger}}]{Saez:2021oxl}%
  \BibitemOpen
  \bibfield  {author} {\bibinfo {author} {\bibfnamefont {B.}~\bibnamefont
  {D\'\i{}az~S\'aez}}, \bibinfo {author} {\bibfnamefont {K.}~\bibnamefont
  {M\"ohling}},\ and\ \bibinfo {author} {\bibfnamefont {D.}~\bibnamefont
  {St\"ockinger}},\ }\bibfield  {title} {\bibinfo {title} {{\rm Two Real Scalar
  WIMP Model in the Assisted Freeze-Out Scenario}},\ }\Eprint
  {https://arxiv.org/abs/2103.17064} {arXiv:2103.17064 [hep-ph]}  (\bibinfo
  {year} {2021})\BibitemShut {NoStop}%
\bibitem [{\citenamefont {Lebedev}(2021)}]{Lebedev:2021xey}%
  \BibitemOpen
  \bibfield  {author} {\bibinfo {author} {\bibfnamefont {O.}~\bibnamefont
  {Lebedev}},\ }\bibfield  {title} {\bibinfo {title} {{The Higgs portal to
  cosmology}},\ }\href {https://doi.org/10.1016/j.ppnp.2021.103881} {\bibfield
  {journal} {\bibinfo  {journal} {Prog. Part. Nucl. Phys.}\ }\textbf {\bibinfo
  {volume} {120}},\ \bibinfo {pages} {103881} (\bibinfo {year} {2021})},\
  \Eprint {https://arxiv.org/abs/2104.03342} {arXiv:2104.03342 [hep-ph]}
  \BibitemShut {NoStop}%
\end{thebibliography}%

\end{document}